# DESIGN AND MODELLING OF DIFFERENT SRAM'S BASED ON CNTFET 32NM TECHNOLOGY

Naagesh. S. Bhat[1]

[1]Developer, Mahindra Satyam Ltd., Bangalore, India
bnsnagesh@gmail.com

## ABSTRACT

*Carbon nanotube field-effect transistor (CNTFET) refers to a field-effect transistor that utilizes a single carbon nanotube or an array of carbon nanotubes as the channel material instead of bulk silicon in the traditional MOSFET structure. Since it was first demonstrated in 1998, there have been tremendous developments in CNTFETs, which promise for an alternative material to replace silicon in future electronics. Carbon nanotubes are promising materials for the nano-scale electron devices such as nanotube FETs for ultra-high density integrated circuits and quantum-effect devices for novel intelligent circuits, which are expected to bring a breakthrough in the present silicon technology.*

*A Static Random Access Memory (SRAM) is designed to plug two needs: i) The SRAM provides as cache memory, communicating between central processing unit and Dynamic Random Access Memory (DRAM). ii) The SRAM technology act as driving force for low power application since SRAM is portable compared to DRAM, and SRAM doesn't require any refresh current. On the basis of acquired knowledge, we present different SRAM's designed for the conventional CNTFET. HSPICE simulations of this circuit using Stanford CNTFET model shows a great improvement in power saving.*

## KEYWORDS

*Carbon nanotube field-effect transistor (CNTFET), Static RAM (SRAM), HSPICE*

## 1. INTRODUCTION

The power consumption has become an important consideration on the VLSI system design and microprocessor as the demand for the portable devices and embedded systems continuously increases [1, 2]. The on-chip caches can reduce the speed gap between the processor and main memory. These on-chip caches are usually implemented using SRAM cells. The write power is usually larger than the read power due to large power dissipation in driving the cell bit lines to full swing. The sum of the power consumption in decoders, bit lines, data lines, sense amplifier, and periphery circuits represents the active power consumption. The power dissipated in bit-lines represents 70 per cent of the total SRAM power consumption during a write operation [3]. Many techniques have been proposed to reduce the write power consumption by reducing the voltage swing level on the bit lines [4-6]. Especially for modern VLSI processor design, SRAM takes large part of power consumption portion and area overhead.

Since the first CNTFET was reported in 1998, great progress has been made during the past years in all the areas of CNTFET science and technology, including materials, devices, and circuits [7]. On the other hand, as the feature size of silicon semiconductor devices scales down to nanometre range, planar bulk CMOS design and fabrication encounter significant challenges [8]. CNTFET among other new materials is promising due to the unique one-dimensional band-structure which reduces backscattering and makes near-ballistic operation. Exceptional electrical properties such as high speed, high-K compatibility, chemical stability, low SCEs have provided CNFETs with

DOI : 10.5121/vlsic.2012.3106        69

excellent characteristics which exceed those of the state of the art Si-based MOSFETs. Several researches have been done to estimate the performance of CNTFET at a single device level in the presence of process related non-idealities and imperfections at the 32 nm technology node using compact CNFET SPICE model [9][10].

While seeking for solutions with higher integration, performance, stability, and lower power, carbon nanotube (CNT) has been presented for next-generation SRAM design as an alternative material in recent years [11]-[15].This paper proposes a novel 4T, 5T, 6T, 7T, 8T, 9T and 10T SRAM cells based on CNTFET to reduce dynamic write-power and to improve the read cycle at the cost of minimal increase of cell area.

## 2. CARBON NANOTUBE FET

Figure 1 illustrates a conceptual layout of a CNT transistor based on Stanford CNFET model. Ideally, several semiconducting CNTs grow on quartz or Si substrate in an exactly straight and parallel pattern. Those segments which are covered by gate are intrinsic CNT regions, whose conductivity is controlled by the gate. Drain and source segments of CNTs are heavily doped to form Ptype or N-type transistor. The drain, gate and source metal contacts and interconnects are defined by conventional lithography. Pitch size, namely the inter-CNT distance, is determined by CNT syntheses process since CNTs are grown in a self-assembly way. Gate width is determined by CNT tube number and pitch.

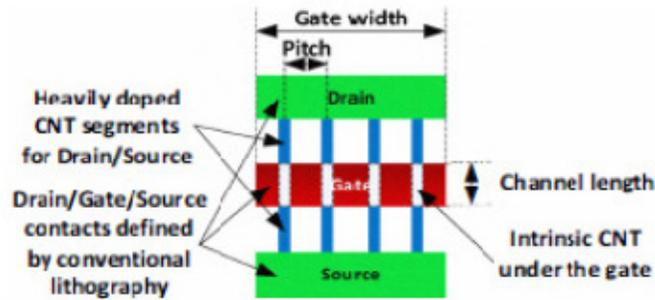

**Figure 1** The CNTFET Layout

CNTFET refers to a field-effect transistor that utilizes a single carbon nanotube or an array of carbon nanotubes as the channel material instead of bulk silicon in the traditional MOSFET structure. It is a three-terminal device consisting of a semiconducting nanotube bringing two contacts (source and drain), and acting as a carrier channel, which is turned on or off electrically via the third contact (gate).

A single-wall carbon nanotube (SWCNT) is a tube formed by rolling a single sheet of graphene. It can either be metallic or semiconducting depends on the chirality vector (m, n), i.e. the direction in that the graphene sheet is rolled. For CNFETs, the threshold voltage of the transistor is defined by the diameter of the carbon nanotubes, which is related to the chirality vector as follows:

$$D_{CNT} = \frac{a}{\pi} \sqrt{m^2 + n^2 + mn}$$

$$V_{TH} = \frac{aV_\pi}{\sqrt{3} * q.D_{CNT}}$$

where q is the charge of an electron, a = 2.49Å is the CNT atomic distance and V$\pi$= 3.033eV is the carbon $\pi$ to $\pi$ bond energy. The sizing of a CNFET is equivalent to adjusting the number of tubes. Since the mobility of n-type and the mobility of p-type carriers inside CNTs are identical, the minimum size is 1 for both P-CNFET and N-CNFET. Semiconducting nanotubes have attracted widespread attention of the electron device and circuit designers as a promising channel material for high-performance transistors. A typical structure of a MOSFET-like CNTFET in planar and co-axial form is illustrated in Figure 2 [16]-[18].

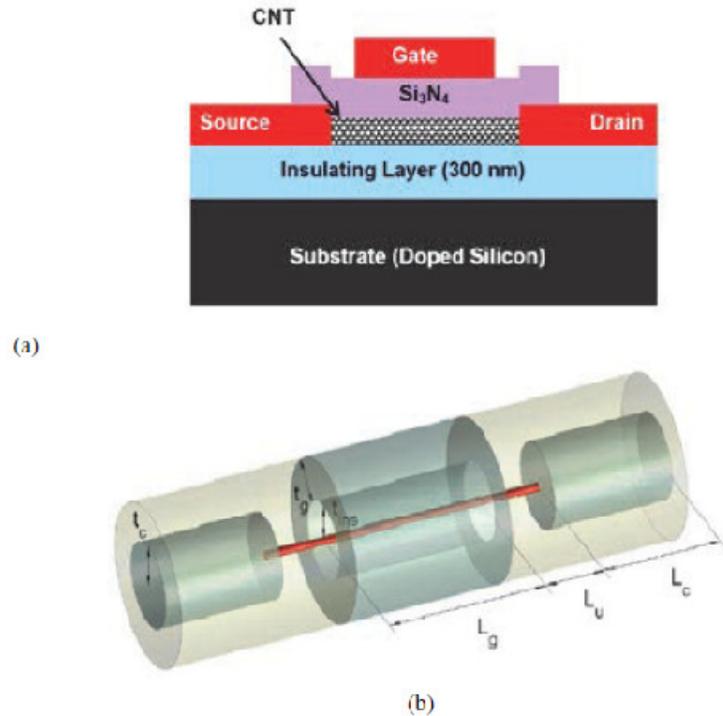

**Figure 0.1** The CNTFET Structures (a) Planar (b) Coaxial

## 3. STATIC RANDOM ACCESS MEMORY (SRAM)

Static Random Access Memory (SRAM) is a type of semiconductor memory. SRAMs are a major component of digital systems such as Embedded systems, microprocessors, reconfigurable hardware, field programmable gate arrays just to name a few. Fast memory access times and design for density have been two of the most important target design criteria for many years, however with device scaling to achieve even faster designs; power supply voltages and device threshold voltages have scaled as well leading to degradation of standby power and static noise margins of memories. SRAM exhibits data remanence, but is still volatile in the conventional sense that data is eventually lost when the memory is not powered. A typical SRAM uses six transistors to store each memory bit. In addition to such 6T SRAM, other kinds of SRAM chips use 4 Transistors till 10 Transistors per bit. The design explains each block of SRAM based on CNTFET 32nm model file.

### 3.1. 4T CNTFET SRAM Cell

Figure 3 shows a circuit equivalent to a developed 4T SRAM cell using a supply voltage of 0.9V. When '0' stored in cell, load and driver transistor are ON and there is feedback between ST node and STB node, therefore ST node pulled to GND by drive transistor and STB node pulled to VDD by load transistor. And when '1' stored in cell, load and driver transistor are OFF and for data retention without refresh cycle following condition must be satisfied. For satisfying the condition when '1' stored in cell, we use leakage current of access transistor, especially sub-threshold current of access transistors. For this purpose during idle mode (when read and write operation don't performed on cell) of cell, BL and BLB maintained at VDD and GND, respectively and word-line1 and wordline2 maintained on VIdle1 and VIdle2, respectively.

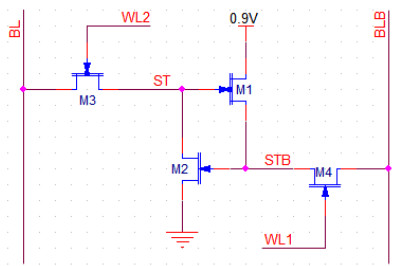

**Figure 3.** New 4T CNTFET SRAM Cell

### 3.1.1. Write Operation

When a write operation is issued the memory cell will go through the following steps.

**Bit-line driving:** For a write, complement of data placed on BLB, and then word-line1 asserted to VDD, but voltages on word-line2 and BL maintained at idle mode (Vword-    line2=VIdle2 and VBL=VDD ).

**Cell flipping:** This step includes two states as follows.

(a) complement of data is zero: in this state, STB node pulled down to GND by   NMOS   access transistor, and therefore the drive transistor will be OFF, and ST  node    will  be  floated  and then pulled up to voltage of BL (VDD) by leakage     current (most   of  this  current  is  sub-threshold current) of PMOS access transistor, and thus load transistor will be OFF.

(b) complement of data is one: in this state, STB node pulled up to VDD-Vtn by   NMOS   access transistor, and therefore the drive transistor will be ON , and ST  node   will   be   pulled   down  to GND, thus load transistor will be ON and STB    node pulled up to VDD.
**Idle mode:** At the end of write operation, cell will go to idle mode and word-line1 and     BLB asserted to VIdle1 and GND respectively.

### 3.1.2. Read Operation

When a read operation is issued the memory cell will go through the following steps.

**Bit-line Pre-charging:** For a read, BL pre-charged to VDD, and then floated. Since, in idle mode BL maintained at VDD, this step didn't include any dynamic energy consumption.

**Word-line activation:** in this step word-line2 asserted to GND and two states can be considered

(a) Voltage of ST node is low: when, voltage of ST node is low, the voltage of BL pulled down to low voltage by PMOS access transistor. We refer to this voltage of BL as VBL-Low.

(b) Voltage of ST node is height: when voltage of ST node is height, the voltage of BL and ST node equalized (we refer to voltage of BL in this state as VBL-High). Since in this state, there is very small different between BL and ST node, dynamic energy consumption is very small.

**Idle mode:** At the end of read operation, cell will go to idle mode and word-line2 and BL asserted to VIdle2 and VDD, respectively.

## 3.2. 5T CNTFET SRAM Cell

In a normal 6T cell both storage nodes are accessed through NMOS pass-transistors. This is necessary for the writing of the cell since none of the internal cell nodes can be pulled up from a stored '0' by a high on the bit line. If this was not the case an accidental write could occur when reading a stored '0'.

However, if the bit lines are not precharged to VCC this is no longer true. With an intermediate precharge voltage, VPC, the cell could be constructed so that a high on the bit line would write a '1' into the cell, but a precharged bit line with a lower voltage would not. Also a low on the bit line could write a '0' into the cell, whereas the intermediate precharge voltage would not, thus giving the cell a precharge voltage window where correct operation is assured. This would eliminate the need for two NMOS transistors, since the cell now can be written both high and low from one side. In turn, that would also result in one less bit line. From a high density point of view this is very attractive. Figure 4 shows the structure of the proposed, resulting five-transistor (5T) SRAM cell. With one less bit line the 5T cell also shares a sense amplifier between two cells. This further reduces the area giving the 5T memory block an even greater advantage over the 6T SRAM.

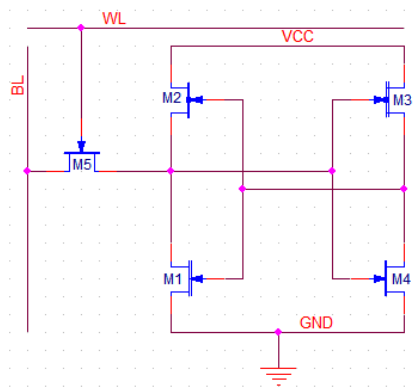

**Figure 4.** 5T CNTFET SRAM Cell

### 3.2.1. Write Operation

Writing in the 5T SRAM cell differs from the 6T cell mainly by the fact that it is done from only one bit line. For the 5T cell the value to be written is held on the bit line, and the word line is asserted. Since the 6T cell was sized so that a '1' could not be written by a high voltage on the bit line, the 5T cell has to be sized differently.

### 3.2.2. Read Operation

The operation scheme when reading a 5T cell is very similar to the 6T SRAM. Before the onset of a read operation, the word line is held low (grounded) and the bit line is precharged. This time however, the bit line is not precharged to *VCC*, but to another value, *VPC*. This value is carefully chosen according to stability and performance requirements. One drawback of the intermediate precharge value is the apparent problem of obtaining this voltage. One obvious way is to supply this voltage externally. The trend today is that microprocessors demand several different supply voltages, so this might in fact not be a significant drawback..

## 3.3. 6T CNTFET SRAM Cell

Each bit in an SRAM is stored on four transistors that form two cross-coupled inverters. This storage cell has two stable states which are used to denote "0" and "1". Two additional access transistors help controlling the access to the cross coupled unit formed by the inverters during read and write operations. So typically it takes six transistors to store one memory bit. The design of a basic SRAM cell is shown in Figure 5. Access to the cell is enabled by the word line (WL) which controls the two access transistors M5 and M6 which allow the access of the memory cell to the bit lines: 'BL' and 'BLbar'. They are used to transfer data for both read and write operations. The presence of dual bit lines i.e. 'BL' and 'BLbar' improves noise margins over a single bit line. The operation of CNFETs based memories is very similar to that of CMOS except for minor differences in device orientation. One such difference being that the source and drain terminals of a CNFET are not interchangeable as is the case with CMOS devices. Care must therefore be taken to orient the transistors in a memory cell in a manner that will ensure correct transmission of logic levels.

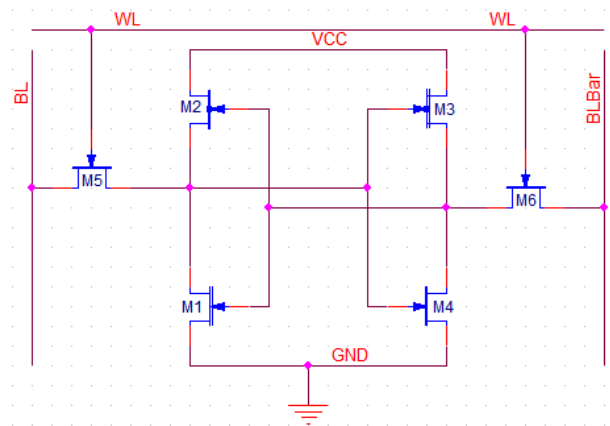

**Figure 5.** 6T CNTFET SRAM Cell

### 3.3.1. Write Operation

The start of a write cycle begins by applying the value to be written and its complement to the bit lines. In order to write a '0', we would apply a '0' to the bit line 'BL' and its complement '1' to the 'BLbar'. A '1' is written by inverting the values of the bit lines i.e. by setting 'BL' to '1' and 'BLbar' to '0'. 'WL' is then made high and the value that is to be stored is latched in. The input drivers of the bit lines are designed to be much stronger than the relatively weak transistors in the cell itself, so that they can easily override the previous state of the cross-coupled inverters. Proper operation of an SRAM cell however needs careful sizing of the transistors in the unit.

### 3.3.2. Read Operation

The read cycle is started by asserting the word line 'WL', enabling both the access transistors M5 and M6. The second step occurs when the values stored in 'Q' and 'Qbar' are transferred to the bit lines 'BL' and 'BLbar' through M1 and M6. On the BL side, the transistors M4 and M5 pull the bit line towards VDD (when a "1" is stored at Q). If the content of the memory was a 0, the reverse would happen and 'BLbar' would be pulled towards 1 and 'BL' towards 0.

### 3.3.3. Idle State

For the idle state, the word line is not asserted and the access transistors M5 and M6 disconnect the cell from the bit lines. The two cross coupled inverters INV1 and INV2 formed by M1, M2 and M3 M4 will continue to reinforce each other as long as they are disconnected from any external circuits.

### 3.4. 7T CNTFET SRAM Cell

The 7-transistor SRAM cell based on CNTFETs has been designed to improve the read cycle and reduce dynamic power. The transistor level schematic of this cell appears in Figure 6. It adds a transistor M7 in the feedback loop and a separate read line 'ReadBit' from the word line 'WriteBit' of the 6-transistor cell.

The four transistors M1, M2 and M3, M4 in the centre form two cross-coupled inverters INV1 and INV2. Due to the feedback structure, a low input value on the first inverter INV1 will generate a high value on the second inverter INV2, which amplifies and stores the low value on the second inverter INV2. Similarly, a high input value on the first inverter INV1 will generate a low input value on the second inverter INV2, which feeds back the high input value onto the first inverter INV1. Therefore, the two inverters INV1 and INV2 will store their current logical value, whatever value that is. But in this circuit feedback connection is established through an extra nMOS transistor M7. The circuit stores data at a node 'Q' and its complement at a node 'Qbar'. This circuit uses two separate transistors M5 and M6 to write and read data from memory cell. To write data into cell 'WriteSelect' signal is used. To read data from the cell 'ReadSelect' signal is used.

This proposed 7T CNTFET SRAM cell depends on cutting off the feedback connection between the two inverters, INV1 and INV2, before a write operation. The feedback connection and disconnection is performed through an extra nMOS transistor M7. During write operation M7 is OFF and during read operation it is ON. The cell depends only on 'WriteBit' to perform a write operation.

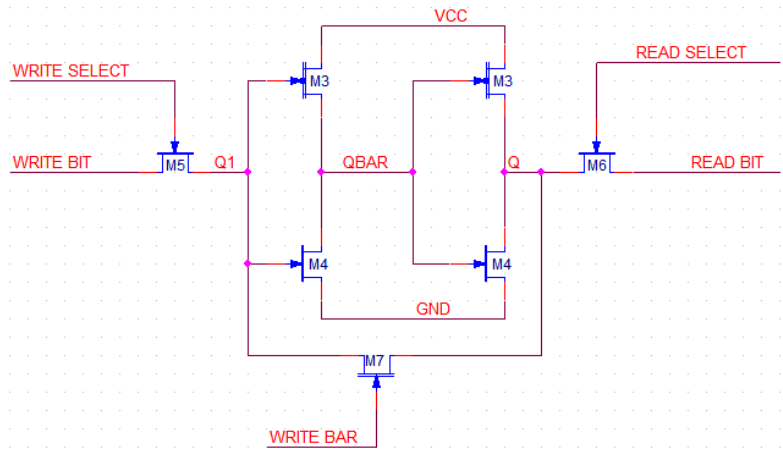

**Figure 6.** 7T CNTFET SRAM Cell

### 3.4.1. Write Operation

The write operation starts by turning M7 off to cut off the feedback connection, thereby allowing for a fast transfer of the logic value from the write bit line 'WriteBit' into the memory cell. 'WriteBit' carries the input data, M5 is turned on by using a signal 'WriteSelect', while M6 is kept off . The 7T SRAM cell looks like two cascaded inverters, INV1 followed by INV2. M5 transistor transfers the data from 'WriteBit' to Q1 which drives INV1, M1 and M2, to develop 'Qbar'. Similarly, 'Qbar' drives INV2, M3 and M4, to develop 'Q', the cell data. Then, M5 is turned off and M7 is turned on to reconnect the feedback link between the two inverters to stably store the new data. Dynamic power reduction would result from the reduced switching activity during memory accesses. The 'WriteBit' line does not have to be pre-charged in preparation for the read operation and a write operation affects only a single bit line of the cell compared to both for the 6-transistor memory cell.

### 3.4.2. Read Operation

Read operation starts by turning on a transistor M6 using a signal 'ReadSelect' and turning off the transistor M5. During this operation feedback path is connected by turning on 'WriteBar' signal. Then the stored data at a node 'Q' can be read at 'ReadBit'. The read cycle is improved based on two aspects of the cell operation namely the ability to pre-charge the read bit line 'ReadBit' irrespective of the activity of the write bit line 'WriteBit' and device sizing of the read zero path with the pull-down transistor M3 of the second inverter made 8 times larger than the M6 to provide a fast path to ground.

### 3.5. 8T CNTFET SRAM Cell

The 8T SRAM cell shown in Figure 7 uses a buffered read to isolate the internal nodes of the cell from the read path. Prior to the read operation the read bit line RBL is precharged to Vdd. The read operation is started by asserting the RWL. RBL either remains at Vdd (if internal node "q" contains a "0") or is pulled down to ground (if internal node "q" contains a "1"). In either case, the internal nodes remain undisturbed. Prior to the write operation, the bit lines are precharged to the pre-determined values. The write operation is initiated by asserting the write word line WWL and the nodes attain the corresponding values from the bit lines.

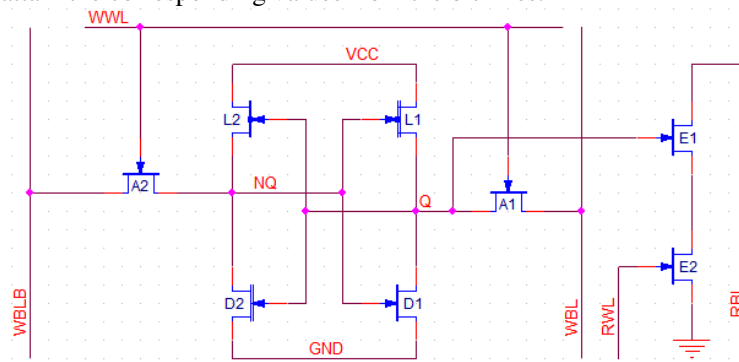

**Figure 7. 8T CNTFET SRAM Cell**

### 3.6. 9T CNTFET SRAM Cell

The proposed 9T SRAM cell consists of cross-coupled inverters formed by the transistors L1, D1, L2 and D2 which store a single bit of information, shown in Figure 8. The write bit line WBL and the pass transistor A2 are used for transferring new data into the cell. Alternatively, the read bit line RBL and transistors E2, E3 and E4 are used for reading data from the cell.

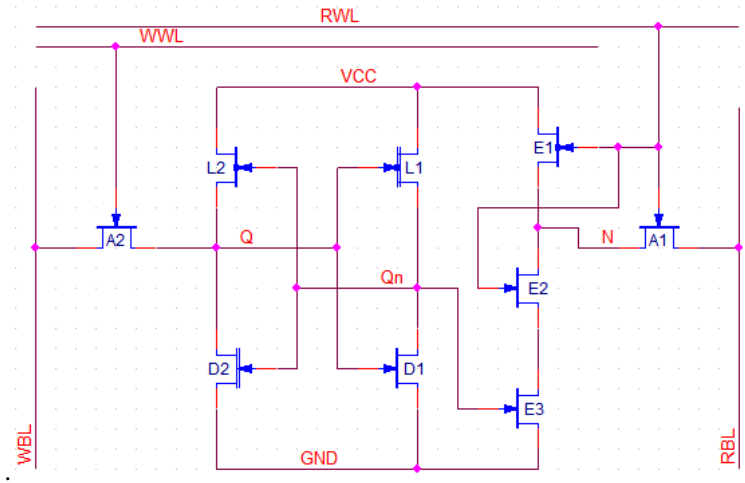

**Figure 8.** 9T CNTFET SRAM Cell

The 9T SRAM cell enhances the read stability by employing a read discharge path that is completely isolated from the internal nodes of the cell. The data stability is thereby significantly improved when compared with the conventional 6T SRAM cell design. Based on the voltage at node "Qb", the RBL is conditionally discharged through the E2-E4 transistor stack during a read operation.

### 3.7. 10T CNTFET SRAM Cell

The 10T SRAM bit cell uses a fully differential read sensing scheme, as shown in Figure 9. In the read mode, WL is enabled and Vgnd is forced to 0 V while WWL remains disabled. The disabled WWL makes data nodes Q and QB decoupled from the bit line during the read access. Due to this isolation, the read SNM of the 10T SRAM cell is almost same as that of the hold SNM of the conventional 6T SRAM cell. Based on the cell data value, one of the bit lines would get discharged after the WL is enabled. It can be noticed that in this 10T SRAM cell, read value is developed as an inverted signal of cell data. Prior to the write operation, the bit lines BL and BLB are precharged to the pre-determined values. In the write mode, both the word lines WL and WWL are enabled to transfer the write data to the cell nodes from the bit lines. Since this 10T SRAM cell has series access transistors, writability is a critical issue.

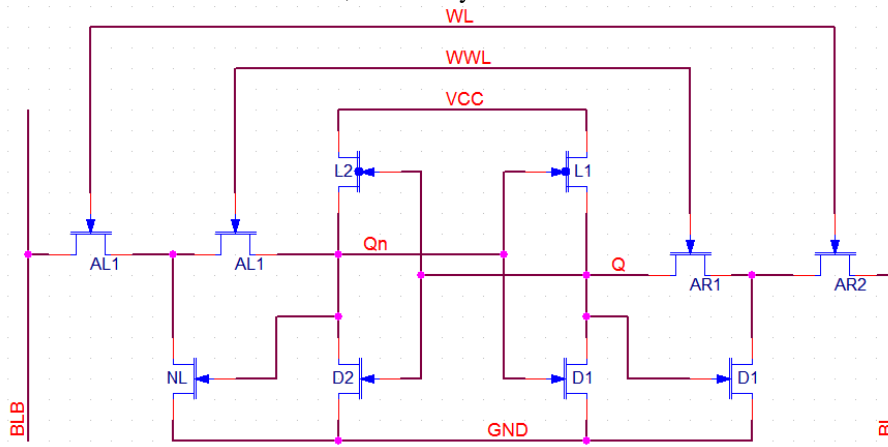

**Figure 9.** 10T CNTFET SRAM Cell

## 4. CNTFET SPECIFICATIONS AND CELL SIZING

The SRAM cells based on CNTFET is designed at 32nm technology. This circuit is simulated in HSPICE using Stanford CNTFET model at 32nm feature size with supply voltage VDD of 0.9V. The following technology parameters are used for simulation of SRAM cells using CNTFET Technology [22-24]:

    Physical channel length (L_channel) = **32.0nm**

    The length of doped CNT source/drain extension region (L_sd) = **32.0nm**

    Fermi level of the doped S/D tube (Efo) = **0.6 eV**

    The thickness of high-k top gate dielectric material (Tox) = **4.0nm**

    Chirality of tube (m, n) = **(19, 0)**

    CNT Pitch = **10nm**

    Flatband voltage for n-CNTFET and p-CNTFET (Vfbn and Vfbp) = **0.0eV and 0.0eV**

    The mean free path in intrinsic CNT (Lceff) = **200.0nm**

    The mean free path in p+/n+ doped CNT = **15.0nm**

    The work function of Source/Drain metal contact = **4.6eV**

    CNT work function = **4.5eV**

The sizing of a CNFET is equivalent to adjusting the number of tubes.

a) **4T CNTFET SRAM Cell**
   a. M1 and M2 Transistors – 3 Tubes
   b. M3 and M4 Transistors – 5 Tubes

b) **5T CNTFET SRAM Cell**
   a. M1 and M4 Transistors – 4 Tubes
   b. M2 and M3 Transistors – 2 Tubes
   c. M5 Transistor – 5 Tubes

c) **6T CNTFET SRAM Cell**
   a. M1, M2, M3 and M4 Transistors – 3 Tubes
   b. M5 and M6 Transistors – 5 Tubes

d) **7T CNTFET SRAM Cell**
   a. M1, M2 and M5 Transistors – 3 Tubes
   b. M4 and M7 Transistors – 1 Tube
   c. M3 Transistors – 8 Tubes
   d. M6 Transistors – 6 Tubes

e) **8T CNTFET SRAM Cell**
   a. L1, L2 and E1 Transistors – 1 Tube
   b. D1 and D2 Transistors – 4 Tubes
   c. A1, A2 and E3 Transistors – 6 Tubes

f) **9T CNTFET SRAM Cell**

    a. E1, E3, E4 and A2 Transistors – 4 Tubes

    b. D1, D2, L1 and L2 Transistors – 1 Tube

    c. E2 Transistor – 7 Tubes

g) **10T CNTFET SRAM Cell**

    a. AL1, AL2, AR1 and AR2 Transistors – 2 Tubes

    b. L1 and L2 Transistors – 3 Tubes

    c. D1 and D2 Transistors – 5 Tubes

    d. NR and NL Transistors – 8 Tubes

## 5. RESULTS AND DISCUSSIONS

All the SRAM cells are designed and verified for successful read, write and hold functionality using CNTFET 32nm technology.

### 5.1. Simulation Waveforms

The simulation waveform of the CNTFET SRAM Cells of 4T SRAM, 6T SRAM and 7T SRAM are shown below in Figure 10, Figure 11 and Figure 12 respectively.

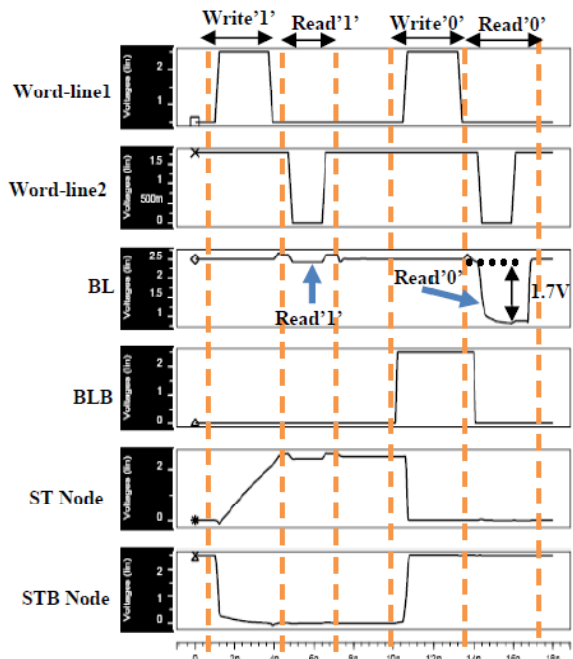

Figure 10. Simulated Waveform of Read/Write Operation of 4T SRAM

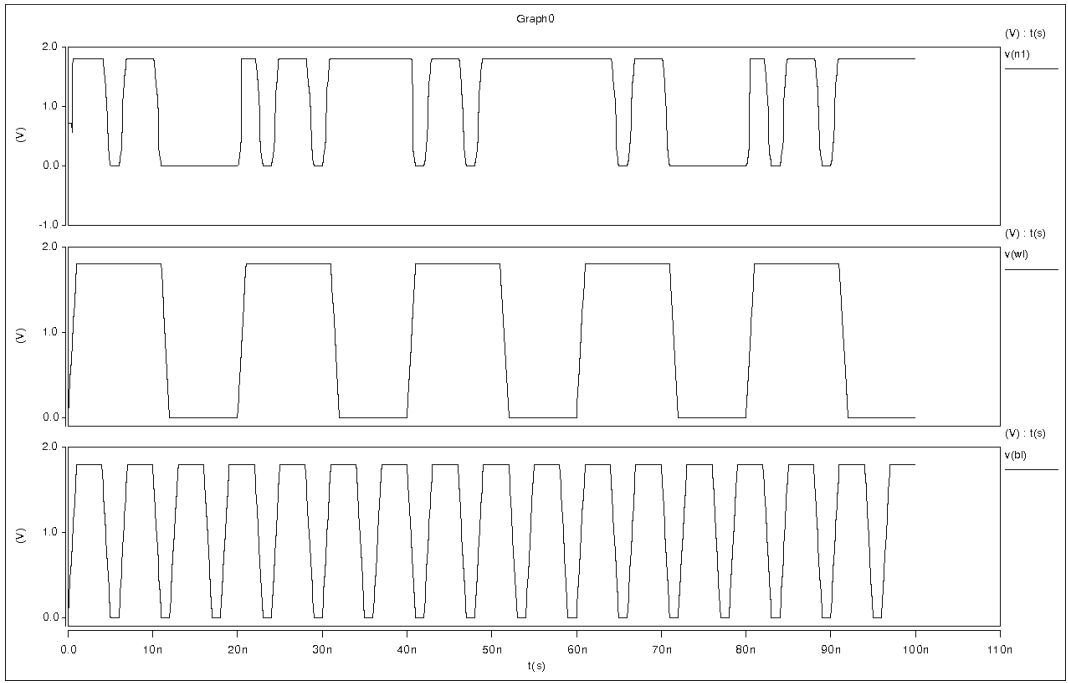

Figure 11. Simulated Waveform of 6T SRAM

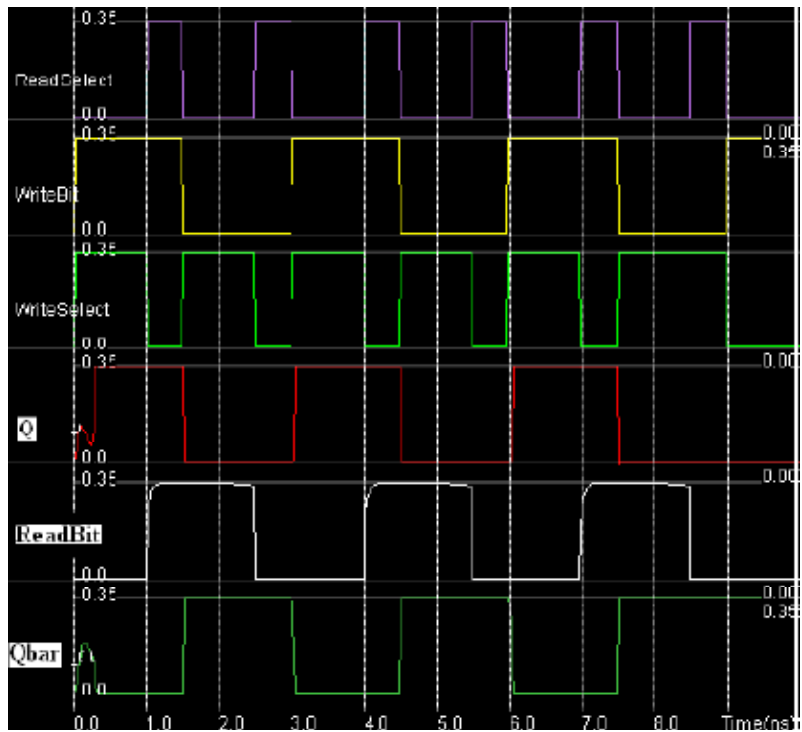

Figure 12. Simulated Waveform of 7T SRAM

## 5.2. Read Static Noise Margin

The read SNMs of the SRAM cells are compared in Table 1. The read SNM of the 8T SRAM, 10 SRAM and 9T SRAM cells is 54% higher when compared to the 6T SRAM cells. The higher read SNM for these cells can be attributed to the fact that they have their internal nodes completely decoupled from the read discharge path.

Table 1. Read Static Noise Margin

| SRAM | Read SNM (uV) |
|---|---|
| 6T SRAM | 195 |
| 8T SRAM | 431 |
| 10T SRAM | 431 |
| 9T SRAM | 431 |

## 5.3. Write Static Noise Margin

The Write Static Noise Margin comparison between the SRAM cells is illustrated in the Table 2. The SRAM cells typically employ write assist technique to boost the writability of the cell and improve the Write Static Noise Margin. Before the onset of the write operation of the 10T SRAM cell, the Vdd floats. This decrease in the supply voltage effectively weakens the SRAM cell making it easy for the access transistors to overpower the node voltages. The 10T SRAM cell boosts the drive voltage of the write access transistors by 30% in order to improve the write performance. The 8T and the 9T SRAM cells do not use write assist techniques and hence have lower write static noise margins than the 10T SRAM cells.

Table 2. Write Static Noise Margin

| SRAM | Write SNM (uV) |
|---|---|
| 6T SRAM | 442 |
| 8T SRAM | 471 |
| 10T SRAM | 571.75 |
| 9T SRAM | 431 |

## 6. CONCLUSION

Carbon-based devices show promising features, so that they are considered as potential candidates to replace silicon based MOSFETs in the future. In this paper a SRAM Cell is designed using CNTFETs at 32nm Technology to reduce write-power dissipation and to reduce the read delay. This circuit is designed and simulated in HSPICE using Stanford CNFET model at 32nm technology. The read SNM of the 8T, 9T and the 10T cells are about 50% higher than the 6T cell. The ease of writability on the 10T SRAM cell is greater than the other cells since it employs a write-assist technique. The 8T, 10T and the 9T cells show a 60% improvement in their mean read SNM values and at least 13% reduction in the standard deviation values. These cells pass the yield criterion with a considerable margin.

## REFERENCES

[1]. B. H. Calhoun, Y. L. Cao, X. Mai, K. L. T. Oileggi, R. A. Rutenbar, and K. L. Shepard, "Digital circuit design challenges and opportunities in the era of nanoscale cmos," Proceedings of the IEEE, vol. 96, no. 2, pp. 343–365, 2007.

[2]. T. Karnik, S. Borkar, and V. De, "Sub-90nm technologies: challenges and opportunities for cad," 2002. 774602 203-206.

[3]. S. Borkar, T. Karnik, S. Narendra, J. Tschanz, A. Keshavarzi, and V. De, "Parameter variations and impact on circuits and microarchitecture," 2003. 775920 338-342.

[4]. S. R. Nassif, "Modeling and analysis of manufacturing variations," in Proceedings of the IEEE Conference on Custom Integrated Circuits, pp. 223–228, 2001.

[5]. M. Orshansky, S. Nassif, and D. Boning, Design for manufacturability and statistical design. Springer Publications, P.O.Box 17, 3300 AA Dordrecht, The Netherlands, 2007.

[6]. S. Nassif, "Delay variability: sources, impacts and trends," in Proceedings of the IEEE International Solid-State Circuits Conference, pp. 368–369, 2000.

[7]. K. A. Bowman, S. G. Duvall, and J. D. Meindl, "Impact of dieto- die and within-die parameter fluctuations on the maximum clock frequency distribution for gigascale integration," IEEE Journal of Solid- State Circuits, vol. 37, no. 2, pp. 183–190, 2002.

[8]. A. Asenov, S. Kaya, and J. H. Davies, "Intrinsic threshold voltage fluctuations in decanano mosfets due to local oxide thickness variations," IEEE Transactions on Electron Devices, vol. 49, no. 1, pp. 112–119, 2002.

[9]. C. C.Chiang and J. Kawa, Design for manufacturability and yield for nano-scale cmos. Springer Publications, P.O.Box 17, 3300 AA Dordrecht, The Netherlands, 2007.

[10]. Semiconductor Industry Association, ITRS Update. San Jose, California, 2008.

[11]. A. J. Bhavnagarwala, T. Xinghai, and J. D. Meindl, "The impact of intrinsic device fluctuations on cmos sram cell stability," IEEE Journal of Solid-State Circuits, vol. 36, no. 4, pp. 658–665, 2001.

[12]. H. Raymond and P. Wang, "Variability in sub-100nm sram designs," in Proceedings of the IEEE International Conference on Computer Aided Design, pp. 347–352, 2004.

[13]. S. Mukhopadhyay, H. Mahmoodi-Meimand, and K. Roy, "Modeling and estimation of failure probability due to parameter variations in nanoscale srams for yield enhancement," in Proceedings of the Symposium on VLSI Circuits, pp. 64–67, 2004.

[14]. L. Chang, D. M. Fried, J. Hergenrother, J. W. Sleight, R. H. Dennard, R. K. Montoye, L. Sekaric, S. J. McNab, A. W. Topol, C. D. Adams, K. W. Guarini, and W. Haensch, "Stable sram cell design for the 32 nm node and beyond," in Proceedings of the Symposium on VLSI Technology, pp. 128–129, 2005.

[15]. C. Benton Highsmith and P. C. Anantha, "A 256-kb 65-nm sub-threshold sram design for ultra-low-voltage operation," IEEE Journal of Solid- State Circuits, vol. 42, no. 3, pp. 680–688, 2007.

[16]. C. I. Joon, K. J. Joon, P. P. Sang, and K. Roy "A 32 kb 10T subthreshold SRAM array with bit-interleaving anddifferential read scheme in 90 nm CMOS," IEEE Journal of Solid-State Circuits, vol. 44, no. 2, pp. 650–659, 2009.

[17]. P. K. Jaydeep and K. Keejong, and K. Roy "A 160 mV, fully differential, robust Schmitt trigger based sub-threshold SRAM," in Proceedings of the International Symposium on Low-Power Electronics, pp. 171–176, 2007.

[18]. S. A. Tawfik and V. Kursun, "Low power and robust 7t dual-vt sram circuit," in IEEE International Symposium on Circuits and Systems, pp. 1452–1455, 2008.

[19]. S. Tavva and D. Kudithipudi, "Variation tolerant 9t sram cell design," in Proceedings of the 20th Great Lakes Symposium on VLSI, pp. 55–60, 2010.

[20]. K. Zhang, Embedded memories for nano-scale vlsis. Springer Publications, 233 Spring Street, New York, NY 10013, 2009.

[21]. S. Tavva and D. Kudithipudi, "Characterization of variation aware nanoscale static random access memory designs," Journal of Low Power Electronics, vol. 6, no. 1, pp. 56–65, 2010.

[22]. J. M.Rabaey, A. Chandrakasan, and B. Nikolic, Digital integrated circuits. Prentice Hall Electronics and VLSI Series, Pearson Education Inc., Upper Saddle River, New Jersey 07458, 2003.

[23]. J. Wang, N. Satyanand, and B. H. Calhoun, "Analyzing static and dynamic write margin for nanoscale srams," in Proceedings of the 13th International Symposium on Low Power Electronics and Design, pp. 129–134, 2008.

[24]. A. J. Bhavnagarwala, S. Kosonocky, C. Radens, K. Stawiasz, R. Mann, Y. Qiuyi and C. Ken, "Fluctuation limits and scaling opportunities for cmos sram cells," in Proceedings of the IEEE International Electronic Devices Meeting, pp. 659–662, 2005.

[25]. P. A. Stolk, H. P. Tuinhoit, R. Duffy, E. Augendre, L. P. Bellefroid, M. J. B. Bolt, J. Croon, C. J. J. Dachs, F. R. Huisman, A. J. Moonen, Y. V. Ponomarev, R. F. M, Roes, R. M. Da, E. Seevinck, K. N. Sreerambhatla, R. Surdeanu, R. M. D. A. Velghe and M. Vertregt, "Cmos device optimization for the mixed-signal technologies," in Proceedings of the International Electronic Devices Meeting, pp. 10.2.1–10.2.4, 2001.

[26]. B. Cheng, S. Roy and A. Asenov, "The scalability of 8t-sram cells under the influence of intrinsic parameter fluctuations," in Proceedings of the 37th European Solid State Device Research Conference, pp. 93– 96, 2007.